\documentclass[aps,prl,superscriptaddress,onecolumn,11pt]{revtex4-2}
\usepackage[utf8]{inputenc}
\usepackage{braket}
\usepackage{qcircuit}
\usepackage{xcolor}
\usepackage{amsmath}
\usepackage{graphicx}
\usepackage{appendix}
\usepackage{comment}
\usepackage{ulem}
\usepackage{physics}
\usepackage{subcaption}

\begin{document}

\title{Universal quantum theory contains twisted logic}

\author{Francesco Atzori}
\affiliation{INRIM, Strada delle Cacce 91, I-10135 Torino, Italy}
\affiliation{Politecnico di Torino, Corso Duca degli Abruzzi 24, I-10129 Torino, Italy}

\author{Enrico Rebufello}
\affiliation{INRIM, Strada delle Cacce 91, I-10135 Torino, Italy}

\author{Maria Violaris}
\affiliation{Clarendon Laboratory, University of Oxford, Parks Road, Oxford OX1 3PU, United Kingdom}
\affiliation{Mathematical Institute, University of Oxford, Woodstock Road, Oxford OX2 6GG, United Kingdom}

\author{Laura T. Knoll}
\affiliation{DEILAP-UNIDEF, CITEDEF-CONICET, J. B. de La Salle 4397, 1603 Villa Martelli, Buenos Aires, Argentina}

\author{Abdulla Alhajri}
\affiliation{Clarendon Laboratory, University of Oxford, Parks Road, Oxford OX1 3PU, United Kingdom}

\author{Alessio Avella}
\author{Marco Gramegna}
\affiliation{INRIM, Strada delle Cacce 91, I-10135 Torino, Italy}

\author{Chiara Marletto}
\author{Vlatko Vedral}
\affiliation{Clarendon Laboratory, University of Oxford, Parks Road, Oxford OX1 3PU, United Kingdom}

\author{Fabrizio Piacentini}
\email{f.piacentini@inrim.it}
\affiliation{INRIM, Strada delle Cacce 91, I-10135 Torino, Italy}

\author{Ivo Pietro Degiovanni}
\author{Marco Genovese}
\affiliation{INRIM, Strada delle Cacce 91, I-10135 Torino, Italy}
\affiliation{INFN, sezione di Torino, via P. Giuria 1, 10125 Torino, Italy}




\begin{abstract}
Quantum theory is notoriously counterintuitive, and yet remains entirely self-consistent when applied universally.
Here we uncover a new manifestation of its unusual consequences.
We demonstrate, theoretically and experimentally (by means of polarization-encoded single-photon qubits), that Heisenberg's uncertainty principle leads to the impossibility of stringing together logical deductions about outcomes of consecutive non-compatible measurements.
This phenomenon resembles the geometry of a Penrose triangle, where each corner is locally consistent while the global structure is impossible.
Besides this, we show how overlooking this non-trivial logical structure leads to the erroneous possibility of distinguishing non-orthogonal states with a single measurement.
\end{abstract}

\maketitle

\section{Introduction}

Unitary, reversible quantum theory has many counter-intuitive consequences. Yet it can be applied with no inconsistencies to the whole universe, including observers and measurement apparatuses; so far, it has also remained unchallenged by experimental evidence. There are also several arguments supporting the universality of quantum theory: for instance, one can argue that once a sector of the universe obeys unitary quantum theory, the rest  must also have non-classical features \cite{dewitt_global_2003, marletto2022quantum}.
These arguments suggest that the emergence of a quasi-classical reality through the mechanism of environmental decoherence \cite{joos_decoherence_2003, zurek_decoherence_2003} or even a Diosi-Penrose-like collapse \cite{diosi_universal_1987, penrose_gravitys_1996, bassi2,bassi1} also requires, at the fundamental level, unitary quantum theory to hold \cite{vedral_observing_2016}.

The consequences of quantum theory's universality have been explored by many thought experiments: starting with Schr\"{o}dinger's cat \cite{schrodinger_collected_2001} and continuing with Everettian quantum theory \cite{everett_foundations_1957, wallace_emergent_2014, vaidman_schizophrenic_1998}. Although alternative interpretations are still possible \cite{int}, it has now become clear that there is no inconsistency between an observer's immediate experience of the classical world and the fact that they obey quantum theory's laws. For instance, an observer may be subject to interference experiments, as explained in a famous proposal by Deutsch \cite{deutsch_quantum_1985, vedral_observing_2016, vedral_observing_2018}, while believing all along to be part of a classical world.

While unitary quantum theory is compatible with an observer's experience of a single world and definite outcomes of measurements, it inevitably forces us to commit to a surprising fact: the real-valued outcomes of measurements are not fundamental, and do not constitute elements of reality -- just like flat spacetime can serve as a local approximation to the real geometry of spacetime, but does not correspond to the reality described by general relativity \cite{deutsch_information_2000}.  In unitary quantum theory, the elements of reality are sets of numbers with a non-commutative algebra, called q-numbers. This non-commutativity reflects the central tenet of quantum information: quantum systems have distinct physical variables that cannot be simultaneously measured to the same arbitrarily high accuracy in a single-shot measurement. As elucidated in  \cite{deutsch_constructor_2015}, this impossibility constraint on measurers of physical variables of given systems (which follows from the no-cloning theorem \cite{wootters_single_1982}  and Heisenberg's uncertainty principle) is the foundation of the rich information-theoretic properties of quantum systems.

In this paper we will provide a novel demonstration of this basic fact: the information-theoretic properties of unitary quantum theory are incompatible with the requirement of there being a single, absolute, sharp outcome of measurements carried out during an interference experiment, where two or more non-commuting observables are relevant.
Specifically, we consider a thought experiment that exposes a novel manifestation of the counterintuitive quantum consequences by identifying a quantum network that represents a sequence of nested measurements of non-commuting observables.
We show that, when the network is analysed by assuming classical logic, we reach a contradiction. Additionally, imposing the classical logic to interpret what is going on leads to the ability to deterministically distinguish non-orthogonal states -- which is tantamaunt to violating Heisenberg's uncertainty principle.
We point out that the network has a logical structure that recalls an impossible geometrical shape known as the `Penrose triangle' \cite{penrose1958impossible}.

Imposing a classical logic structure on quantum thought experiments leads to a number of alleged paradoxes, that have been erroneously attributed to quantum theory being inconsistent \cite{hardy_1993, frauchiger_quantum_2018}. These paradoxes simply arise from assuming a set of supplementary classical axioms in addition to unitary quantum theory, which violate its unitary structure.  Such axioms are not at all needed to make sense of a quasi-classical reality within unitary quantum theory. Unitary quantum theory is perfectly consistent with itself, as well as with the experience of fleeting, but definite, relative observed outcomes of experiments.
In the relative-state picture, these outcomes give rise to a quasi-classical reality whose domain of applicability is limited both in space and time.
Indeed, post-diction on the state of a quantum system is generally not possible, with its use leading to contradictions and paradoxes \cite{hardy_quantum_1992,pig,pig2,Vai13}

We suggest that this construction could provide a geometrical interpretation of the quantum branch structure, as suggested also in \cite{isham_possible_2000}.  Our result vindicates the idea that assuming unitary quantum theory to describe a portion of an experiment (with the potential re-interference of different branches of the quantum wave-function)  is logically incompatible with also assuming that the outcomes of measurements in different branches are persistent and absolute.

\section{Results}

\subsection{Quantum Penrose triangle network} \label{triangle}

\begin{figure}
\centering
\includegraphics[scale=1.5]{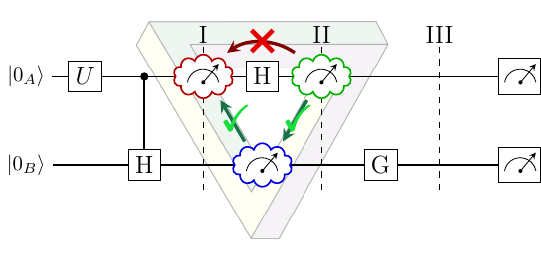}
\caption{\textbf{Quantum circuit demonstrating the impossibility of using classical logic on measurement outcomes.} Regions I, II and III are indicated by dashed lines. While H is used to indicate Hadamard gates, $U$ and $\mathrm{G}$ represent, respectively, the unitary evolution and generalized Hadamard-like rotation in Eq. \eqref{G}.
The Quantum Penrose Triangle structure appears clearer when considering the case $\theta=\frac{\pi}{2}$, i.e. for $\mathrm{G}=\mathrm{H}$; by measuring qubit A to be $\ket{1}$ (green cloud), we can conclude that qubit B is $\ket{1}$ in region II (blue cloud); but, if qubit B is $\ket{1}$ in region II, then qubit A must have been $\ket{1}$ in region I (red cloud), i.e. before the Hadamard gate.
This seems to lead us to the paradoxical situation in which we have perfect knowledge of the state of qubit A both before and after H, although we know that this is not possible, since the Hadamard gate does not commute with z-basis measurements.}
\label{original short}
\end{figure}

Here we present an experiment whereby it appears that (misleading) inductive reasoning maps onto the logical structure of a Penrose triangle.
We will use the quantum circuit in Figure \ref{original short}.
%
%
%
A two-qubit state $\ket{0_A0_B}$ propagates through this circuit, where the H's indicate Hadamard gates and $U$ and G are:
\begin{equation}
U = \begin{bmatrix}
\alpha & \;\;-\beta^* \\
\\
\beta & \;\;\alpha^*
\end{bmatrix}
\;\;\;\;\;\;
G = \begin{bmatrix}
\cos\left(\frac{\theta}{2}\right) & \sin\left(\frac{\theta}{2}\right) \\
\\
\sin\left(\frac{\theta}{2}\right) & -\cos\left(\frac{\theta}{2}\right)
\end{bmatrix}
\label{G}
\end{equation}
with $|\alpha|^2+|\beta|^2=1.$
Initially, qubit A undergoes the unitary evolution induced by $U$.
A Controlled-H gate acts afterwards on the two-qubit state, followed by a H gate on qubit A between regions I and II and the $\mathrm{G}$ gate on qubit B between regions II and III, before the two qubits are measured.
The global state moves through the following stages (subscripts omitted for the sake of readability):
\begin{align}
    & \ket{00} \nonumber\\ \label{line 2}
    & \Rightarrow \alpha \ket{00} + \beta \ket{10} \\ \label{line 3}
    & \Rightarrow\; \mathrm{I}:\;\; \alpha \ket{00} + \frac{\beta}{\sqrt{2}}\ket{10} + \frac{\beta}{\sqrt{2}}\ket{11} \\ \label{line 4}
    & \Rightarrow\; \mathrm{II}:\;\;  \left(\frac{\alpha}{\sqrt{2}}+\frac{\beta}{2}\right)\ket{00} + \left(\frac{\alpha}{\sqrt{2}}-\frac{\beta}{2}\right) \ket{10} + \frac{\beta}{2}\ket{01} - \frac{\beta}{2}\ket{11}\\ \label{line 5}
    & \Rightarrow\; \mathrm{III}:\;\;  \left[c\frac{\alpha}{\sqrt{2}}+(c+s)\frac{\beta}{2}\right]\ket{00} + \left[c\frac{\alpha}{\sqrt{2}}-(c+s)\frac{\beta}{2}\right] \ket{10} + \left[s\frac{\alpha}{\sqrt{2}}+(s-c)\frac{\beta}{2}\right] \ket{01}+ \left[s\frac{\alpha}{\sqrt{2}}+(c-s)\frac{\beta}{2}\right] \ket{11}
\end{align}
where we have used the notation $c = \cos\left(\frac{\theta}{2}\right)$ and $s = \sin\left(\frac{\theta}{2}\right)$.
Now, let us restrict to the scenario with $\theta=\frac{\pi}{2}$, i.e. with $\mathrm{G}=\mathrm{H}$.
Consider the case where the final measurement of qubit A gives $\ket{1}$, thus reducing Eq. \eqref{line 4} to having two terms, $\ket{10}$ and $\ket{11}$.
If the coefficient of one of these terms is $0$, then we can deduce with certainty whether qubit B is in the $\ket{0}$ or in the $\ket{1}$ state.
Setting the coefficient of $\ket{11}$ to $0$ means $\beta = 0$, which would lead us to the trivial case where the global state is $\ket{00}$ until each qubit is measured in the H basis.
However, we could alternatively set $U\rightarrow U_+:\;\frac{\alpha}{\sqrt{2}}-\frac{\beta}{2} = 0$, such that $\alpha = \sqrt{\frac13}$ and $\beta = \sqrt{\frac23}$, deducing with certainty that qubit B in region II must be in the $\ket{1}$ state upon measuring qubit A in the $\ket{1}$ state.
Now, given that qubit B is certainly in the $\ket{1}$ state in II, we can conclude from Eq. \eqref{line 3} that, before the measurement of qubit A in the Hadamard basis, qubit A was for certain in the $\ket{1}$ state (region I).
This suggests that, stepping back to Eq. \eqref{line 2}, only the $\ket{10}$ term could exist (given our final measurement of qubit A as $\ket{1}$).
Then, the Controlled-H gate must have definitely applied a Hadamard to qubit B, and the measurement of qubit B in region III will certainly give $\ket{0}$ as an outcome, meaning that the state $\ket{11}$ is ``forbidden'' at the end of the circuit.\\
Following the same logic for the case where the measurement of qubit A gives $\ket{0}$, we instead set $U\rightarrow U_-:\;\frac{\alpha}{\sqrt{2}}+\frac{\beta}{2} = 0$ to find $\alpha = \sqrt{\frac13}$ and $\beta = -\sqrt{\frac23}$.
This leads to the same conclusion that, at the end of the circuit, the measurement of qubit B will for certain give $\ket{0}$ as an outcome at the end of the circuit (given that $\mathrm{HH}=\mathcal{I}$, being $\mathcal{I}$ the identity operator), and the $\ket{01}$ state is now forbidden in region III. \\
Taking the case in which the final measurement of qubit A gives either $\ket{1}$ or $\ket{0}$, depending on the unitary evolution $U_\pm$ initially applied to $\ket{0_A}$, through a series of logical deductions we argued that the state of qubit A in I must have been certainly $\ket{1}$ (since the Controlled-H gate has acted on qubit B).
This claim is clearly faulty: we cannot draw a conclusion about the state of qubit A in I, because the H gate between regions I and II does not commute with the measurement, which is in the computational basis. \\
The deduction that qubit A was in the state $\ket{1}$ $(\ket{0})$ in the past can seem reasonable, because it stems from two deductions that are each individually consistent.
To see this, consider the set of deductions needed to conclude qubit A's past state in the case $U_+$ ($U_-$) has been applied to qubit B, represented by arrows in Fig. \ref{original short}: \\
(1). If qubit A is measured in the state $\ket{1}$ ($\ket{0}$) at the end of the circuit, considering a unitary evolution $U=U_+$ ($U=U_-$), then qubit B must be in the state $\ket{1}$ in region II. \\
(2). If qubit B is in the state $\ket{1}$ in region II, then qubit A must be in the state $\ket{1}$ before the H gate (region I). \\
(3). Therefore, if qubit A is measured in the state $\ket{1}$ ($\ket{0}$) at the end of the circuit with $U=U_+$ ($U=U_-$), then qubit A was in the state $\ket{1}$ in region I. \\
Statement (1) is true: if qubit A is measured and gives $\ket{1}$ ($\ket{0}$), then a measurement of qubit B in the computational basis in region II would indeed have $\ket{1}$ as outcome.
Statement (2) is also true: if a measurement of qubit B in II would give $\ket{1}$ as outcome in the computational basis, then a measurement of qubit A in I would also yield the state $\ket{1}$.
The problem is that these logical deductions cannot be chained together: we cannot conclude that, if qubit A is measured to be $\ket{1}$ $(\ket{0})$ in III, then qubit B must have been in the state $\ket{1}$ in II, and \emph{therefore} qubit A must have been in the state $\ket{1}$ in I.
Indeed, both statements refer to the state of qubit A but in different bases (given that a H gate acts on qubit A between regions I and II): having a defined state of qubit A both before and after a H gate would violate the uncertainty principle, since we would simultaneously know the precise state of a qubit measured in two mutually unbiased bases.\\
We therefore find that quantum theory sustains a logical structure which resembles that of a Penrose triangle, as depicted in Fig. \ref{original short}.
The corners of the Penrose triangle are each consistent individually, but when chained together the global triangle is impossible; similarly, the two logical deductions about the state of one qubit based on the other one are individually consistent, but impossible when chained together.\\
Another paradoxical outcome rising from applying classical logic to this scenario is the (apparent) possibility of discriminating non-orthogonal states with $100\%$ accuracy.
Indeed, the line of reasoning reported above can be resumed by saying that, when implementing the unitary evolution $U_+$ on qubit A, then a measurement of this qubit yielding $\ket{1}$ at the end of the circuit would imply, for $\mathrm{G}=\mathrm{H}$, qubit B to be $\ket{0}$.
Similarly, when $U_-$ occurs, qubit A resulting in the $\ket0$ state would yield the measurement outcome $\ket1$ for qubit B.
Hence, if qubit A initial evolution is restricted to $U=U_\pm$, when the result for qubit B measurement is $\ket{0}$ then we can determine from the measurement of qubit A whether the unitary $U$ prepared the state $\ket*{\psi_{U_+}}=\sqrt{\frac{1}{3}}\ket{0} + \sqrt{\frac{2}{3}}\ket{1}$ (qubit A resulting to be $\ket1$) or $\ket*{\psi_{U_-}}=\sqrt{\frac{1}{3}}\ket{0} - \sqrt{\frac{2}{3}}\ket{1}$ (qubit A resulting to be $\ket0$).\\
This ability to deterministically discriminate these two non-orthogonal states contradicts the conclusions we can draw from Eq. \eqref{line 5}, where if qubit B is measured in $\ket{0}$ then there is the possibility of qubit A being $\ket{0}$ or $\ket{1}$, although with different probabilities.
Therefore, while the seemingly logical deductions we made from Eq. \eqref{line 4} would seem to enable us to fully distinguish two non-orthogonal states of qubit A, Eq. \eqref{line 5} indicates that such a discrimination is possible up to a certain error threshold, given by the Helstrom bound \cite{helb}, proportional to the overlap between the two states under consideration.

%
%
%
\subsection{Experimental demonstration}

\begin{figure}
\centering
\includegraphics[scale=0.11]{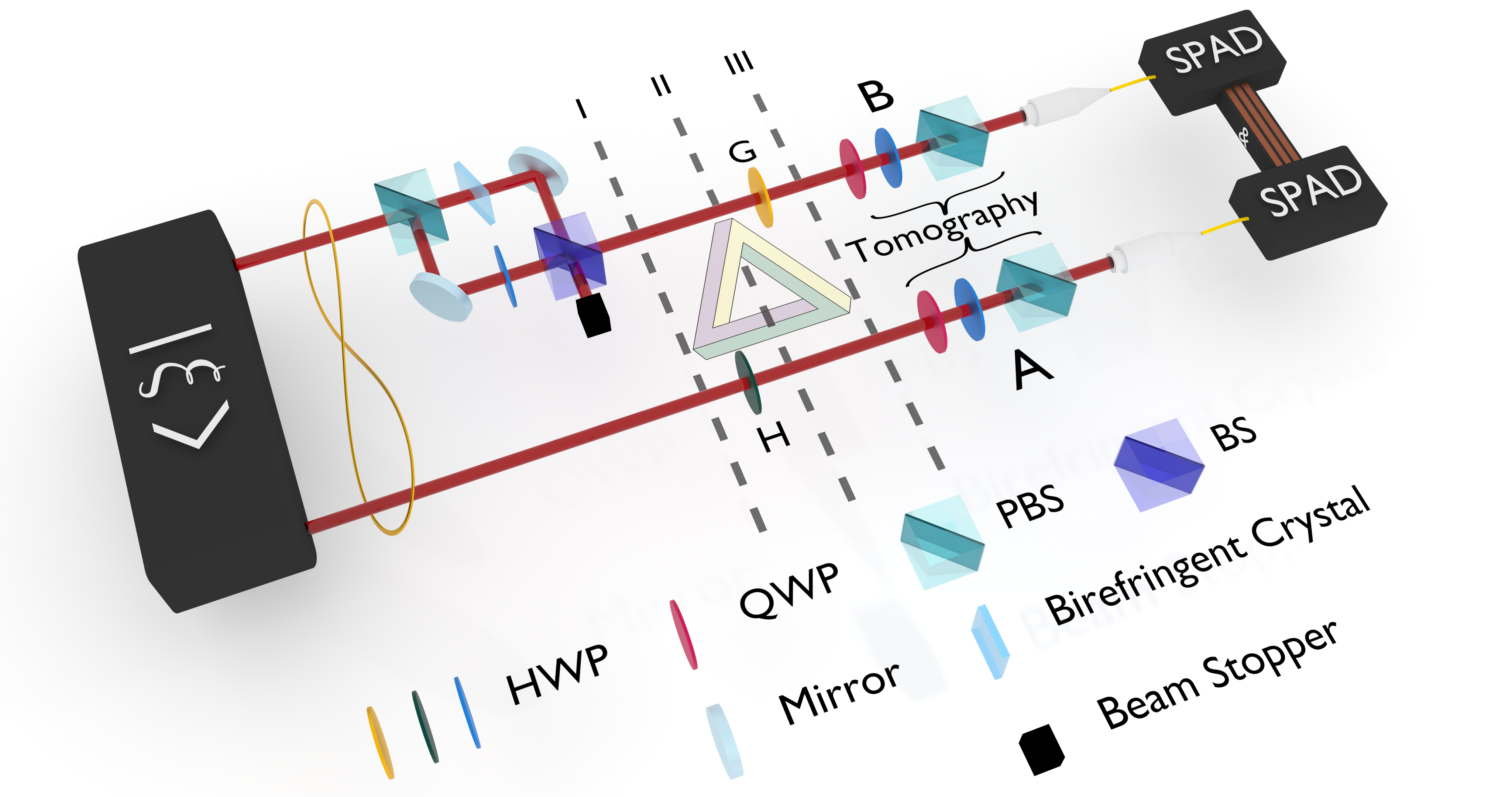}
\caption{\textbf{Pictorial representation of the experimental scheme.}
To realize the $\ket{\psi_{I+}}$ state of Eq. \eqref{psi1}, first our polarization-entangled photon pairs source (black box) is tuned to produce the state $\ket{\xi}=\sqrt{\frac{1}{3}} \ket{0_A 0_B} + \sqrt{\frac{2}{3}}\ket{1_A 1_B}$, then photon B enters a Mach-Zehnder-like interferometer with an input PBS, a HWP and a birefringent crystal hosted in the two paths, and a 50:50 recombination BS of which only one face output is collected.
The HWPs $\mathrm{H}$ and $\mathrm{G}$, mounted on two rails, are inserted time-by-time to realize the corresponding gates and obtain the states needed in regions I, II and III of the quantum circuit in Fig. \ref{original short}.
Two sets (one per branch) composed of a QWP, a HWP, and a PBS allow performing tomographic reconstruction of the two-photon quantum state, detected by two SPADs connected to a time-tagger used to record coincidence events between them.
HWP: Half-wave plate; QWP: Quarter-wave plate; PBS: Polarizing beam splitter; BS: $50:50$ non-polarizing beam splitter; SPAD: Single-photon avalanche diode.}
\label{app}
\end{figure}

\begin{figure}[t]
\centering
\begin{subfigure}{.3\textwidth}
    \centering
\caption{$\rho_{\psi_{I+}}^{\mathrm{th}}$}
    \includegraphics[width=\textwidth]{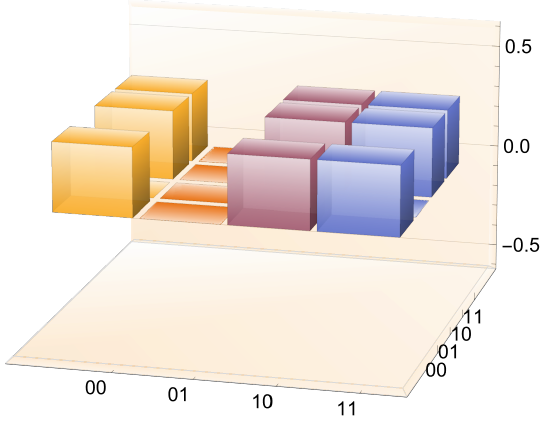}
\label{tomoUpPsi1th}
\end{subfigure}%
\begin{subfigure}{.3\textwidth}
    \centering
\caption{$\mathrm{Re}\left[\rho_{\psi_{I+}}^{\mathrm{rec}}\right]$}
    \includegraphics[width=\textwidth]{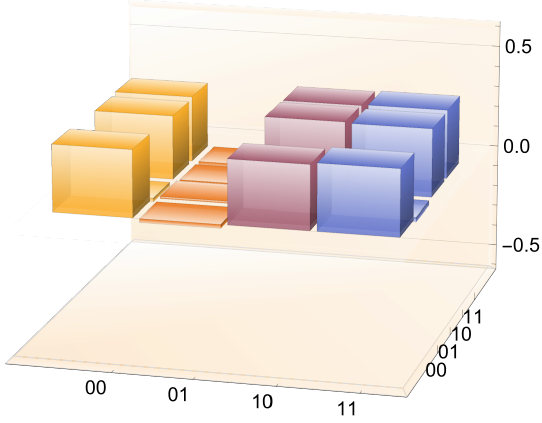}
\label{tomoUpRePsi1rec}
\end{subfigure}
\begin{subfigure}{.3\textwidth}
    \centering
\caption{$\mathrm{Im}\left[\rho_{\psi_{I+}}^{\mathrm{rec}}\right]$}
    \includegraphics[width=\textwidth]{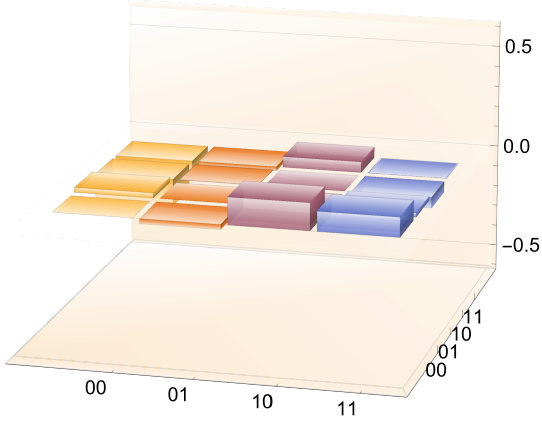}
\label{tomoUpImPsi1rec}
\end{subfigure}\\%
%
%
\begin{subfigure}{.3\textwidth}
    \centering
\caption{$\rho_{\psi_{II+}}^{\mathrm{th}}$}
    \includegraphics[width=\textwidth]{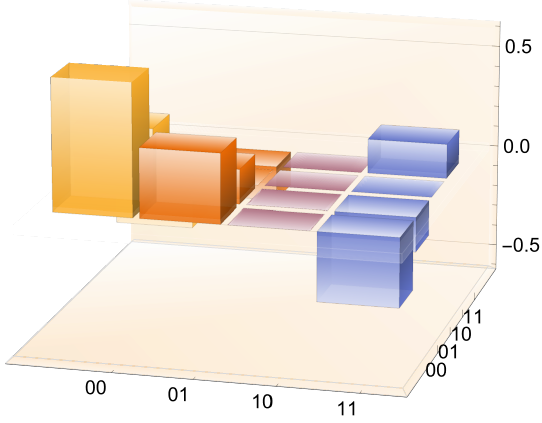}
\label{tomoUpPsi2th}
\end{subfigure}%
\begin{subfigure}{.3\textwidth}
    \centering
\caption{$\mathrm{Re}\left[\rho_{\psi_{II+}}^{\mathrm{rec}}\right]$}
    \includegraphics[width=\textwidth]{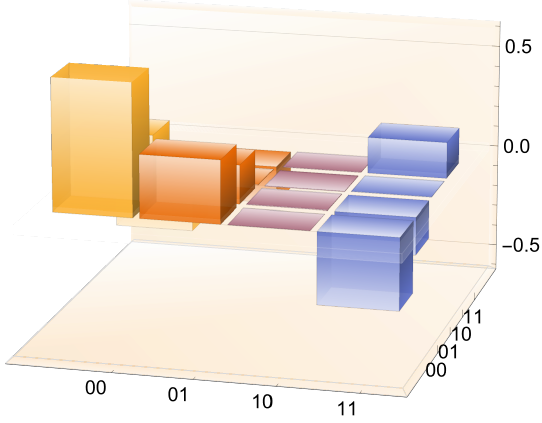}
\label{tomoUpRePsi2rec}
\end{subfigure}
\begin{subfigure}{.3\textwidth}
    \centering
\caption{$\mathrm{Im}\left[\rho_{\psi_{II+}}^{\mathrm{rec}}\right]$}
    \includegraphics[width=\textwidth]{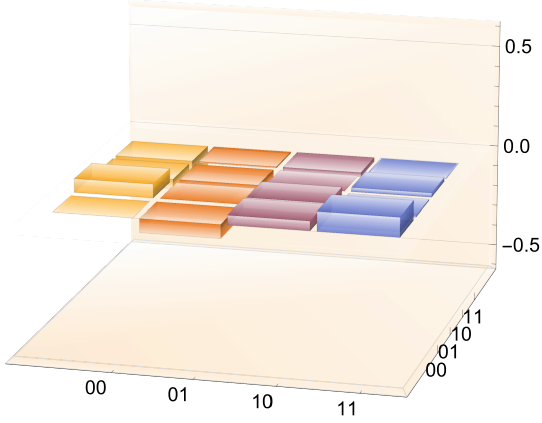}
\label{tomoUpImPsi2rec}
\end{subfigure}\\%
%
%
%
\begin{subfigure}{.3\textwidth}
    \centering
\caption{$\rho_{\psi_{III+}}^{\mathrm{th}}$}
    \includegraphics[width=\textwidth]{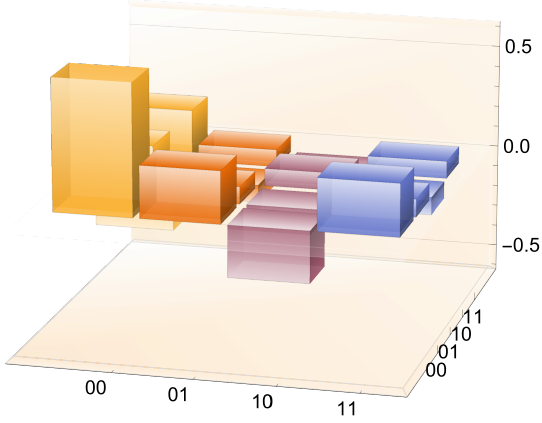}
\label{tomoUpPsi3th}
\end{subfigure}%
\begin{subfigure}{.3\textwidth}
    \centering
\caption{$\mathrm{Re}\left[\rho_{\psi_{III+}}^{\mathrm{rec}}\right]$}
    \includegraphics[width=\textwidth]{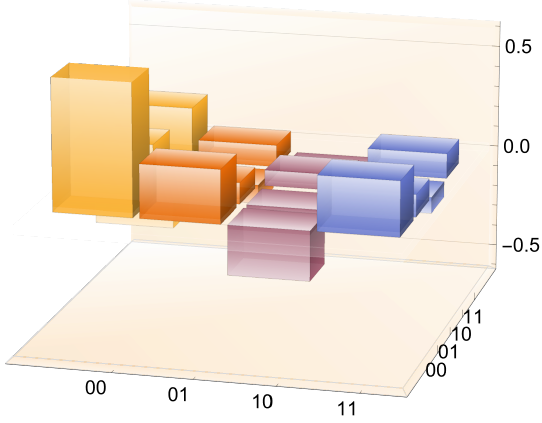}
\label{tomoUpRePsi3rec}
\end{subfigure}
\begin{subfigure}{.3\textwidth}
    \centering
\caption{$\mathrm{Im}\left[\rho_{\psi_{III+}}^{\mathrm{rec}}\right]$}
    \includegraphics[width=\textwidth]{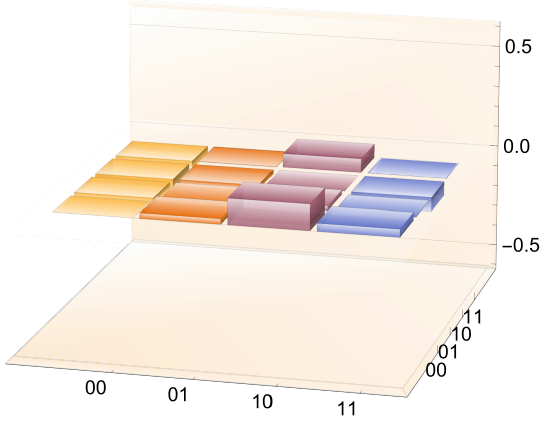}
\label{tomoUpImPsi3rec}
\end{subfigure}\\%
\caption{\textbf{Quantum tomography of the two-qubit state in regions I-III}. Quantum tomographic reconstructions (indicated by the ``rec'' label) of the density matrices of the quantum states generated in regions I, II and III (for $\theta=\frac{\pi}{2}$, i.e. with G=H) of our setup, compared with their theoretical counterparts (indicated by the ``th'' label) given by Eqs. \eqref{psi1}-\eqref{psi3+} (the real part is the only one reported, since the imaginary one is identically 0). Results obtained for $U=U_-$ are reported in the Supplementary Information.}
\label{varGU}
\end{figure}

We can demonstrate the ``Quantum Penrose Triangle'' structure stemming from classical inferential logic by experimentally implementing the quantum circuit in Fig. \ref{original short}, showing that statements (1) and (2) in the previous section are true by verifying the global quantum states in Eq.s \eqref{line 4} and \eqref{line 3}, respectively.
Furthermore, by verifying the finite probability of measuring $\ket{11}$ in Eq. \eqref{line 5} we show that drawing post-dictions on the photon state from statement (3) leads to a contradiction with quantum theory.
For the experimental demonstration we consider $\theta=\{0,\frac{\pi}{4},\frac{\pi}{2}\}$ for the $\mathrm{G}$ gate, thus gradually changing the basis of qubit B measurement in region III from the z-basis to the x-basis.
Specifically, for $U=U_+$, the states in regions I, II and III are (see Supplementary Information for the case $U=U_-$):
\begin{align}
\ket{\Psi_{I+}} & = \sqrt{\frac{1}{3}}\ket{00} + \frac{1}{\sqrt{3}}\ket{01} + \frac{1}{\sqrt{3}}\ket{11}; \label{psi1}\\
\ket{\Psi_{II+}} & = \frac{1}{\sqrt{6}}(2\ket{00} + \ket{01} - \ket{11}); \label{psi2+}\\
\ket{\Psi_{III+}} & = \frac{1}{\sqrt{6}}\left((2c+s)\ket{00} + (2s-c)\ket{01} - s\ket{10} + c\ket{11}\right). \label{psi3+}
\end{align}
We can now experimentally determine how the outcomes predicted by the classical logic compare to that of quantum theory, exploiting the experimental setup schematized in Fig. \ref{app} (see Methods for more details).
The creation of the states in Eq. \eqref{psi1} requires the implementation of a Controlled-H gate.
To realize it, polarization-entangled photon pairs at $\lambda=$810 nm are generated via degenerate type-II Spontaneous Parametric Down Conversion (SPDC), and prepared in the state $\ket{\xi}=\sqrt{\frac{1}{3}} \ket{0_A 0_B} + \sqrt{\frac{2}{3}}\ket{1_A 1_B}$, being $0$ and $1$ the horizontal and vertical polarizations respectively.
Then, photon B enters a Mach-Zehnder-like interferometer with a polarizing beam splitter (PBS) as input port and a $50:50$ non-polarizing beam splitter (BS) as output port.
Within the interferometer, a half-wave plate (HWP) rotates the $\ket{1_B}$ component to $\ket{+_B}=\sqrt{\frac{1}{2}}(\ket{0_B}+\ket{1_B})$.
A fine adjustment of the length mismatch $\Delta l$ between the two arms of the interferometer generates a relative phase $\exp(i \pi \Delta l/\lambda)$ in $\ket{\xi}$, allowing to achieve the equivalent of the unitary evolution operator $U_+$.
After recombining the two arms of the interferometer in the final BS, and discarding one of the BS facets output, the bipartite state reads $\ket{\psi_{I+}}=\sqrt{\frac{1}{3}}(\ket{0_A 0_B}+\ket{1_A 0_B}+\ket{1_A 1_B})$.
The $\mathrm{H}$ and $\mathrm{G}$ gates in Fig. \ref{original short} are implemented using two HWPs.
Afterwards, the state characterization is carried out by means of two (one per branch) quantum tomographic apparata \cite{Bog10} composed of a quarter-wave plate (QWP), a HWP, and a PBS.
Finally, photons A and B are fiber-coupled and detected by two single-photon avalanche diodes (SPADs) connected to a time-tagger.
By inserting time-by-time the two HWPs realizing the quantum gates $\mathrm{H}$ and $\mathrm{G}$, one can produce and reconstruct the states $\ket{\psi_{I+}}$ , $\ket{\psi_{II+}}$, and $\ket{\psi_{III+}}$ of Eqs. \eqref{psi1}-\eqref{psi3+}, that can be observed in correspondence of the three dashed lines in Fig. \ref{app}.\\
Experimental results are reported in Fig. \ref{varGU}, 
where the reconstructed density matrices $\rho_\psi^{\mathrm{rec}}$ (being $\psi$ a label indicating the target state) are compared with the theoretically-expected ones, $\rho_\psi^{\mathrm{th}}$.
The Uhlmann Fidelity $F=\left[\Tr\left(\sqrt{\sqrt{\rho_\psi^{\mathrm{th}}}\rho_\psi^{\mathrm{rec}}\sqrt{\rho_\psi^{\mathrm{th}}}}\right)\right]^2$ \cite{fid} is evaluated for each reconstructed state.
The fidelity values obtained, all above $F\geq0.97$, demonstrate the high quality of the states produced in each of the three regions of the setup.
\begin{figure}[t]
\centering
    \includegraphics[width=0.8\textwidth]{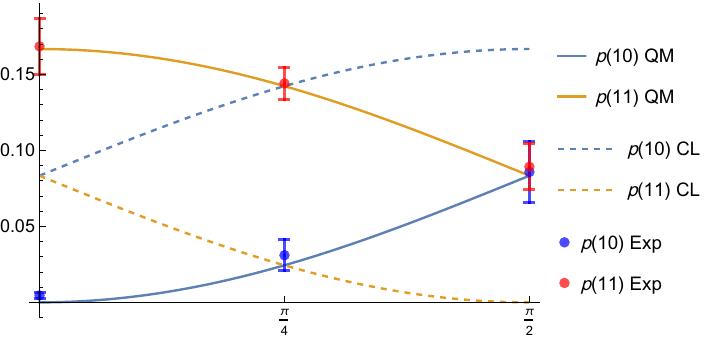}
\caption{\textbf{Measurement of the classically ``forbidden'' states.} Experimental probability of measuring the states $\ket{1_A1_B}$ and $\ket{1_A0_B}$ (``$p(11)$ Exp'' and ``$p(10)$ Exp'', respectively red and blue dots, together with their uncertainty bars) at the end of the circuit given the initial unitary evolution $U=U_+$, compared with the corresponding quantum mechanical predictions (``$p(11)$ QM'' and ``$p(10)$ QM'', respectively yellow and blue solid curve) and the ones given by classical logic (``$p(11)$ CL'' and ``$p(10)$ CL'', respectively yellow and blue dashed curve), as a function of the $\mathrm{G}$ gate parameter $\theta$.
Following classical inductive reasoning, observation of the state $\ket{1_A1_B}$ should be forbidden for $\theta=\frac{\pi}{2}$, and only $\ket{1_A0_B}$ should be present.
Results obtained for $U=U_-$ are reported in the Supplementary Information.}
\label{varG}
\end{figure}
Additionally, in order to show how the classical inductive logic leads to faulty conclusions when applied to the circuit in Fig. \ref{original short}, it is evaluated the probability $p^{i_A j_B}_+(\theta)$ of measuring the state $\ket{i_A j_B}$ in III for $U=U_+$ ($i,j=0,1$).
Indeed, classical reasoning would imply that the state $\ket{1_A 1_B}$ should be forbidden in III for $U=U_+$ when $\theta= \frac{\pi}{2}$ (i.e., when $\mathrm{G}=\mathrm{H}$), and only its ``counterpart'' $\ket{1_A 0_B}$ should appear.
Conversely, by looking at the experimental data (red dots and bars, representing the experimental uncertainties), we can see how this does not happen, since the classical predictions (yellow and blue dashed lines for the forbidden and allowed states, respectively) are in strong disagreement with the experimental observations, perfectly matched by the quantum mechanical predictions (yellow and blue solid lines for the forbidden and allowed states, respectively).
%
\section{Discussion}

Our Penrose triangle structure demonstrates how using classical logic in a quantum setting leads to contradictions.
Classically, we are able to refer to the past state of a given system after its future state has been measured, whereas quantum theory prevents us from combining logical statements that refer to the same qubit at different points in time when it has been manipulated by operations which do not commute with the measurement.
Our logic is very general, as it underlies several other distinctive quantum features arising from consecutive measurements of non-commuting observables (e.g. Leggett-Garg inequalities \cite{leggett1985quantum}, entanglement in time \cite{brukner2004quantum}, contextuality \cite{kochen1990problem}, non-locality \cite{prep,vir24}, and recent methods of treating space and time on an equal footing in quantum theory \cite{fitzsimons2015quantum,PDO1,PDO2}).\\
Our setup emphasises how the notion of ``branch structure'' in unitary quantum theory can be misleading.
In expositions of unitary quantum theory, when a qubit enters a superposition relative to some basis, it can be described as two ``branches''.
When it becomes entangled with another qubit, the qubits are correlated in each branch.
However, a key point that can be overlooked is that the branches are able to interfere, and hence the splitting of branches can change as a quantum circuit evolves.
Specifically, in our example, the impossibility of post-dictions about qubit A in the z-basis before a H gate cannot be made based on the outcome of measuring qubit A in the z-basis after such a gate, as there is not a single quantum ``branch'' connecting the past state with the measurement outcome.
This occurs because the H gate does not commute with the z-basis measurement.
Therefore we reinforce that a necessary consequence of unitary quantum theory is that branches are able to interfere.\\
The subtlety in why the deductions cannot be strung together also has a role in recently claimed paradoxes in quantum theory.
As mentioned earlier, the quantum network and classical logic imposed follows similar reasoning to the Frauchiger-Renner paradox, which claims that quantum theory cannot consistently describe observers within quantum theory \cite{frauchiger_quantum_2018}.
The paradox can be directly mapped onto our circuit by interpreting the quantum gates as being implemented by observers, who make deductions about other observers based on their results.
The reasoning is also related to other thought experiments purported to show paradoxical or surprising features of quantum theory, including Hardy's paradox \cite{hardy_quantum_1992} and the pigeonhole paradox \cite{pig,pig2}. \\
We have drawn a parallel between deductions that can be made within unitary quantum theory and a geometrical structure, which raises the question as to how far we can interpret counterintuitive structures within quantum theory geometrically, as has been suggested in other quantum contexts \cite{isham_possible_2000}.
A similar link between local possibility and global impossibility arising due to the non-commutativity of observables has been identified for local elements of reality, when investigating the nonlocality of a photon \cite{heaney_extreme_2011}.
In this work, the local elements of reality are mapped onto a Penrose square.
Additionally, there has been a comparison to Penrose-triangle structure in a work on contextuality \cite{abramsky_contextuality_2020}.
Further investigation into how non-commutativity in quantum theory leads to impossible geometries when using classical logic could unify these examples of this property of quantum theory, which could form the basis of a future investigation.\\
We have discussed a mechanism by which it naively appears that we can distinguish non-orthogonal quantum states better than quantum theory allows, leading to an internal contradiction in quantum theory.
We have demonstrated that this contradiction arises from applying classical logic to quantum deductions, and chaining together incompatible deductions that are individually consistent.
The logical structure we have identified resembles the geometry of a Penrose triangle, with locally consistent and globally inconsistent statements paralleling the locally consistent corners of a triangle impossibly chained together.
Understanding this twisted quantum logic could give insights into quantum branching structure, help navigate apparent paradoxes, and link with other manifestations of impossible geometries arising from the non-commutativity of operators in quantum theory.
This is further substantiated by our experimental demonstration of this effect using high-quality single-photon qubits.
On this basis, we reiterate the appeal to take unitary quantum theory seriously,  rather than marveling at the fact that it defies naive classical expectations. This defiance of naive expectations is proper of all most fundamental theories of nature (from Newton's laws to special and general relativity), and quantum theory is no exception.
Whether we will be able to find a better theory, that improves on quantum theory itself, is conditional on us finally accepting the full set of consequences of unitary quantum theory and its counter-intuitive, but fully consistent, account of our quasi-classical experiences.

\section*{Methods}

\subsection*{Experimental setup details}
\begin{figure}
\centering
\includegraphics[scale=0.5]{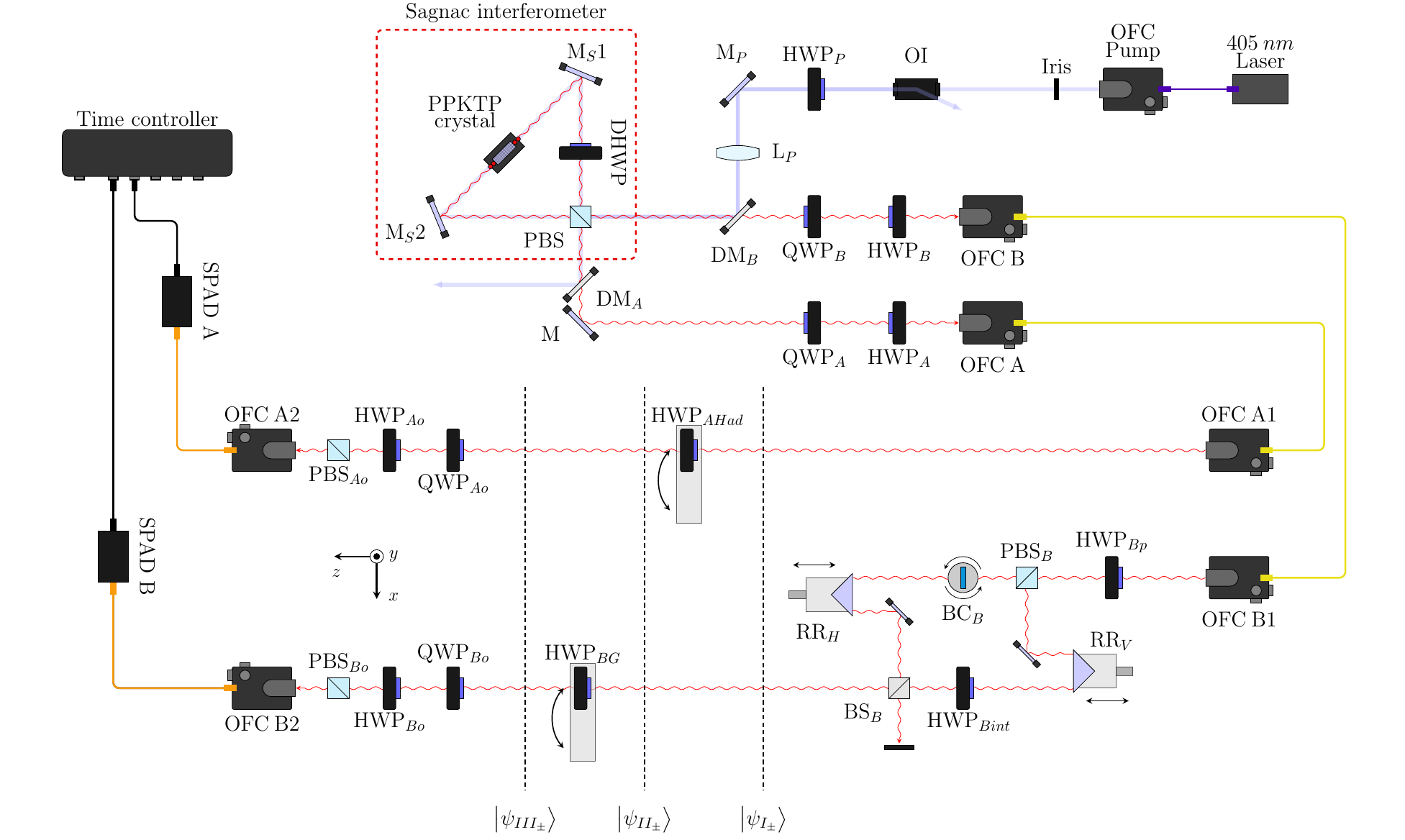}
\caption{\textbf{Detailed experimental setup.}
Polarization-entangled photon pairs are generated in a Sagnac interferometer, with each photon coupled to an optical fiber and sent to one of the two branches.
The Mach-Zehnder-like interferometer allows adjusting the relative phase of the state and rotating only one of the polarization components of one photon oh the pair, gives the possibility to realize one of the two evolutions $U_\pm$.
Two HWPs mounted on two rails allow generating the three states needed.
Afterwards, two (QWP+HWP+PBS) sets, one for each photon, enable the tomographic reconstruction of the two-photon quantum state.
After the PBSs, photons  are fiber-coupled and addressed to two silicon single-photon avalanche diodes with associated time-tagging electronics.
OFC: Optical fiber coupler; L: lens; OI: Optical isolator; HWP: Half-wave plate; QWP: Quarter-wave plate; DHWP: Dual half-wave plate; DM: Dichroic mirror; M: Mirror; PBS: Polarizing beam splitter; BC: Birefringent crystal; BS: 50:50 non-polarizing beam splitter; RR: Retroreflector; PPKTP: Periodically-poled Potassium titanyl phosphate; SPAD: Single photon avalanche diode.}
\label{expapp}
\end{figure}

A detailed scheme of our experimental setup is shown in Fig. \ref{expapp}.
Orthogonally-polarized photons at 810 nm are generated via degenerate collinear type-II spontaneous parametric down-conversion (SPDC) occurring in a Sagnac interferometer that hosts a periodically-poled potassium titanyl-phosphate (PPKTP) crystal pumped by a 405 nm CW laser.
Along the pump path, the L$_p$ lens, paired with the objective of the optical fiber coupler (OFC pump), enables the generation of a Gaussian beam with its waist in correspondence of the center of the PPKTP crystal.
The optical isolator (OI) prevents potential back-reflections to the laser source and prepares the pump photons in a horizontal linear polarization state, which can be eventually rotated by a half-wave plate (HWP$_p$).
A dichroic mirror (DM$_B$), highly reflective for the pump and highly transmissive for the down-converted photons, sends the pump laser to the polarizing beam splitter (PBS).
Here, the horizontal and vertical polarization components of the pump are sent to the PPKTP crystal, traveling clockwise and counterclockwise in the Sagnac interferometer, respectively.
A dual-wavelength half-wave plate (DHWP), with an angle of 45$^\circ$ with respect to the horizontal plane, ensures that the photons entering the PPKTP crystal both clockwise and counterclockwise are horizontally-polarized, generating type-II SPDC.
Consequently, the Sagnac interferometer structure allows obtaining a coherent superposition of the two paths, resulting in the two down-converted photons exiting the interferometer in the state $\ket{\xi' (\vartheta)} = \cos (\vartheta) \ket{H_AV_B} + e^{i \varphi} \sin (\vartheta) \ket{V_AH_B}$ where $\vartheta$ represents the pump polarization rotation angle (with respect to the horizontal plane) induced by HWP$_p$, $\varphi$ represents a generic phase adjustable by changing the position of the PPKTP crystal, and A and B are the two branches where the photons are addressed to before being spectrally filtered and coupled to a single-mode fiber.
The dichroic mirrors (DM$_{A(B)}$) prevent residual pump transmission into the down-converted photons channels.
Subsequently, the combination of quarter- and half-wave plates (QWP$_{A(B)}$ and HWP$_{A(B)}$, respectively) compensates for unwanted polarization changes in the single-mode fibers.
To prepare the state in Eq. \eqref{psi1}, photon B first passes through HWP$_{B_p}$, rotating $H \leftrightarrow V$, to achieve the state $\ket{\xi (\vartheta)} = \cos (\vartheta) \ket{H_AH_B} + e^{i \varphi} \sin (\vartheta) \ket{V_AV_B}$.
Afterwards, photon B enters a Mach-Zehnder-like interferometer composed of a polarizing beam splitter (PBS$_B$) at the entrance and a $50:50$ non-polarizing beam splitter (BS$_B$) at the exit.
The path length of the two branches can be adjusted using two retroreflector mirrors mounted on two translators with micrometric drivers (RR$_H$ and RR$_V$) to control the phase in the interferometer.
This phase can be fine-tuned by rotating the birefringent crystal (BC$_B$), allowing to chose between the evolution operators $U_\pm$.
The half-wave plate (HWP$_{Bint}$), set at a $\frac{3\pi}{8}$ angle with respect to the horizontal plane, rotates the vertical component of photon B to diagonal polarization ($\ket{V_B} \rightarrow \sqrt{\frac{1}{2}}(\ket{H_B}+\ket{V_B})$).
Then, the two wave function components of photon B interfere at BS$_B$, and only one output port is selected.
By tuning the total relative phase $\varphi+ \pi \Delta l/\lambda$ and by rotating the HWP$_p$ to approximately $27.37^\circ$, corresponding to $\cos(\vartheta)\simeq0.577\simeq\sqrt{\frac13}$, the state $\ket{\psi_{I+}}=\sqrt{\frac{1}{3}}(\ket{H_A H_B}+\ket{V_A H_B}\pm\ket{V_A V_B})$ is prepared.
The sets of QWP$_{A_o(B_o)}$+HWP$_{A_o(B_o)}$+PBS$_{A_o(B_o)}$ are utilized for the tomographic reconstruction of the density matrix of the bipartite state.
The photons are subsequently collected, coupled to a multimode fiber, and detected by a Single-Photon Avalanche Diode (SPAD$_{A(B)}$) on each branch.
A time-tagger registers coincidence detection events.
Finally, the two half-wave plates, HWP$_{AHad}$ and HWP$_{BG}$, mounted on two rails, can be time-by-time inserted to implement, respectively, the $\mathrm H$ and $\mathrm G$ gates, in order to prepare the quantum states $\ket{\psi_{II+}}$ and $\ket{\psi_{III+}}$ (or the corresponding states for the $U_-$ evolution).
Specifically, when neither HWP$_{AHad}$ nor HWP$_{BG}$ is inserted, $\ket{\psi_{I+}}$ is obtained; inserting only HWP$_{AHad}$, rotated of a $\frac{\pi}{8}$ angle with respect to the horizontal plane, allows for the reconstruction of $\ket{\psi_{II+}}$; finally, with both HWP$_{AHad}$ and HWP$_{BG}$ inserted, $\ket{\psi_{III+}}$ is obtained.
By varying the rotation angle of HWP$_{BG}$, the parameter $\theta$ of the G gate is changed, allowing to obtain the data set shown in Fig. \ref{varG}.

\section{Acknowledgements}
This work was financially supported by the projects QuteNoise (call ``Trapezio'' of Fondazione San Paolo) and AQuTE (MUR, call ``PRIN 2022'', grant No. 2022RATBS4). This work was also funded by the project Qu-Test, which had received funding from the European Union's Horizon Europe under the grant agreement number 101113901.

\section{Author contributions}
The work in the lab was mainly performed by F.A. and E.R, in collaboration with A.Av., M.Gr. L.T.K. and F.P., who supervised the activity.
The idea stemmed from M.V. and C.M., and was developed by M.V. (that, together with F.A. and E.R., is the first author of this paper) and A.Al. under the supervision of C.M. and V.V.
The experiment was planned by I.P.D. and F.P, with the support of M.Ge.
The project was supervised by V.V. and C.M. for the theoretical part and, for the experimental one, by I.P.D. and M.Ge., leading the quantum optics sector of INRIM. They, together with M.Gr. and F.P. are also the leaders of the projects funding this work. The paper was initially drafted by M.V. and F.P., with contributions from all authors to the final version.

\section{Competing interests}
The authors declare no competing interests.

\section{Materials \& Correspondence}
Correspondence and materials requests should be sent to: Fabrizio Piacentini (f.piacentini@inrim.it).

\end{document}